# Low Temperature Magnetic Studies on $PbFe_{0.5}Nb_{0.5}O_3$ Multiferroic

Shidaling Matteppanavar[1], Basavaraj Angadi[1,*], Sudhindra Rayaprol[2]

[1] Department of Physics, Bangalore University, Bangalore –560056

[2] UGC-DAE-CSR, Mumbai Centre, R-5 Shed, B.A.R.C, Mumbai – 400085

**Abstract**

The $PbFe_{0.5}Nb_{0.5}O_3$ (PFN), a well-known $A(B'_{1/2}B''_{1/2})O_3$ type multiferroic was successfully synthesized in single phase by a single step solid state reaction method. The single phase PFN was characterized through XRD, microstructure through SEM, and magnetic studies were carried out through temperature dependent vibrating sample magnetometer (VSM) and neutron diffraction (ND) measurements. PFN exhibits a cusp at around 150 K in the temperature dependent magnetic susceptibility corresponding to the Néel temperature ($T_{N1}$) and another peak around 10 K ($T_{N2}$) corresponding to spin-glass like transition. In the temperature dependent ND studies, a magnetic Bragg peak appears at Q = 1.35 Å$^{-1}$ (where Q = $4\pi\sin\theta/\lambda$, is called the scattering vector) below $T_N$ (150 K) implying antiferromagnetic (AFM) ordering in the system. On the basis of Rietveld analysis of the ND data at $T$ = 2 K, the magnetic structure of PFN could be explained by a G-type antiferromagnetic structure.

1. **INTRODUCTION**

   $PbFe_{0.5}Nb_{0.5}O_3$ (PFN) is a well-known $A(B'_{1/2}B''_{1/2})O_3$ type multiferroic [1], which has received much attention in recent years due to its potential for technological applications. An intense research is underway on PFN due to the unique magneto-electric (magnetic moment and electric dipole moment) coupling behavior, which makes it very appealing from both the theoretical and the technological points of view [1-3].

Lead based multiferroic systems, PbFe$_{0.5}$Nb$_{0.5}$O$_3$ (PFN), PbFe$_{0.5}$Ta$_{0.5}$O$_3$ (PFT), and PbFe$_{2/3}$W$_{1/3}$O$_3$ (PFW) have attracted attention because in these compounds, the magnetic; Fe$^{3+}$ *(d$^n$)* ions and nonmagnetic *(d$^0$)*; Nb$^{5+}$, Ta$^{5+}$, and W$^{6+}$ ions share the B$^I$ and B$^{II}$ site of the A(B$^I$ B$^{II}$)O$_3$ perovskite, respectively. The *(d$^5$)* ion in the BO$_6$ octahedral site leads to ferromagnetic order while the *(d$^0$)* ions at the same lattice position provide ferroelectric (FE) order [3–5].

PbFe$_{0.5}$Nb$_{0.5}$O$_3$ (PFN) was considered to be ferroelectrically and antiferromagnetically ordered below its Néel temperature ($T_N$ ~ 145 K) [6]. PFN undergoes transition from paraelectric (PE) to ferroelectric (FE) at Curie temperature of around 385 K [7-8], due to the structural transition from centro-symmetric (cubic) to non-centrosymmetric (monoclinic), which is an indication of an existence of ferroelectricity in the system. In addition to ferroelectric and antiferromagnetic features, the magnetoelectric (ME) coupling observed below $T_N$ offers great interest in PFN for potential applications as well as for fundamental understanding. Though there are few reports on the study of nuclear and magnetic structure in PFN, due to the various different structures reported at different temperature intervals, no conclusive structure model was achieved yet. Lampis et al. [2] through neutron diffraction and X-ray diffraction studies reported that PFN has monoclinic structure (*Cm*) below 250 K, Cubic (*Pm3m*) above 376 K and in between it is in tetragonal (*P4mm*). Ivanov et al. [9] through neutron diffraction showed the rhombohedral structure with *R3c* symmetry at both 300 K and 10 K. Bonny V et. al. [10] from their X-ray and synchrotron data on single crystals suggested a small monoclinic distortion away from rhombohedral symmetry at room temperatures, together with the existence of an intermediate phase of tetragonal symmetry at temperatures between 355K and the cubic structure above the ferroelectric Curie point of 376 K. Hence, the structure of PFN is under debate and is a subject of study.

Single phase formation of the Pb based compounds largely depends upon the synthesis conditions, however, due to magnetic anisotropy or inhomogeneity or competing interactions, a system might exhibit phase coexistence. While synthesizing Pb based (such as PFN) compounds, one must consider the loss of lead oxide and has to be careful about the reactivity of each component in the material. That is, two factors play a major role in the synthesis of any lead based complex perovskite viz., (a) high volatility of PbO as its evaporation temperature is around 750 °C and (b) High reactivity of B″ oxide over B′ with PbO in Pb(B′B″)O$_3$. The PbO evaporation results in vacancies at Pb and O sites and also leads to the formation of secondary phases in event of deficiency of lead [11, 12]. This result in the formation of $A_2B_2O_7$ based pyrochlore phases as parasite secondary phase during the synthesis of PFN. The formation of secondary phases has considerable effect on the physical properties of the resultant material (PFN).

Therefore it is necessary to optimize the synthesis parameters and conditions to produce single phase PFN for better understanding of structure and properties for possible practical applications. In this article we report the synthesis method for producing single phased PFN by optimizing the synthesis conditions by using the modified solid state reaction route. We also present and discuss the low temperature ND and magnetization data of PFN for its structure and magnetic properties.

2. **EXPERIMENTAL DETAILS**

*Single-step method (modified solid state reaction method)*: AR grades of Pb(NO$_3$)$_2$, Fe$_2$O$_3$ and Nb$_2$O$_5$ were taken in stoichiometric quantities and ground in pestle and mortar using ethanol medium for 2 hours. The resultant powder was subjected to calcination at 650 °C/2 hrs. After calcination, the powder was ground again with polyvinyl alcohol (PVA) as binder.

The dried powder was then cold-pressed uni-axially at 50 kN pressure and then sintered at 1050 $^o$C for 1 hour in a closed Pb rich environment to minimize the PbO evaporation [13, 14].

The sintered samples were characterized by X-ray diffraction (XRD; *Phillips 1070*) using Cu-K$_\alpha$ radiation (wavelength, $\lambda$ = 1.5406 Å), for phase purity. Scanning Electron Microscopy (SEM) was used for studying the microstructure and morphology. Magnetization studies were carried out on a vibrating sample magnetometer (VSM) attached to a physical property measurement system (*Quantum Design* PPMS). Neutron diffraction measurements were carried out on a focusing crystal based powder diffractometer, available at UGC-DAE-CSR beam line in Dhruva reactor, BARC. Neutrons at a wavelength 1.48 Å were used for the present study. Rietveld analysis were carried out on powder XRD and ND data using the *Fullprof* suite programs for crystallographic as well as for magnetic structure studies [15].

3. RESULTS AND DISCUSSIONS

*3.1 Phase and microstructural analysis*

Figure 1 shows XRD pattern of PFN obtained by single step solid state reaction method. Using the structural model given in Ref [3], the XRD data was refined using the Rietveld method. The analysis confirms that the sample prepared, using the method described above, is in single phase and has no secondary parasite phases like $Pb_2Nb_2O_7$ or $Pb_2Nb_4O_{13}$.

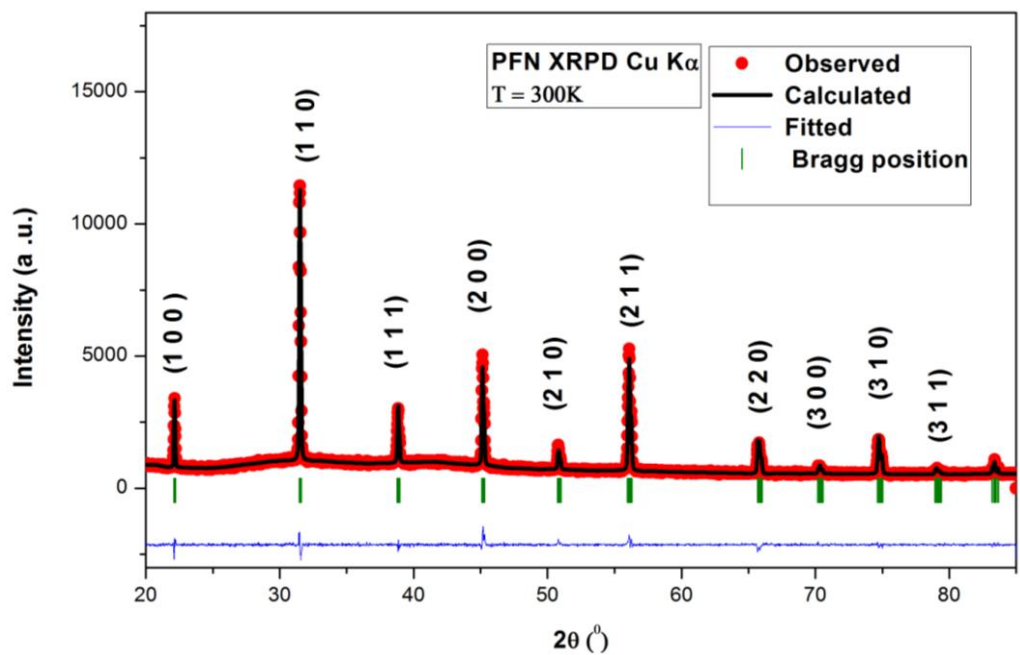

**Figure 1.** Room temperature X-ray powder diffraction data of PFN. The Rietveld analysis was carried out assuming a monoclinic structure.

The low temperature calcination and sintering proved to be effective in achieving the single phase without the secondary phases. Good agreement is found between observed and calculated profiles for XRD data. The SEM micrograph of PFN is shown in Fig. 2. From the micrograph, average grain size is found to be 2 μm.

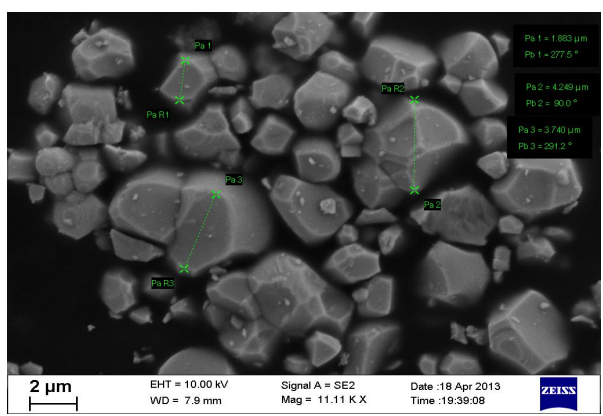

**Figure 2.** SEM micrograph of PFN sintered at 1150 °C for 1 hour exhibiting surface microstructure

*3.2 Magnetization studies*

Magnetic susceptibility ($\chi = M/H$) of PFN measured in both zero field cooled (ZFC) and field cooled (FC) in a field of 500 Oe is shown in Fig. 3. The ZFC curve increases monotonically on decreasing temperature from 300 K, and exhibits a cusp around 150 K ($T_{N1}$), indicating the onset of antiferromagnetic ordering. However on further decrease in temperature, $\chi$ increases further and exhibits another peak around 10 K ($T_{N2}$) before falling rapidly as $T$ approaches to 5 K. The FC curve also show similar features at $T > 10$ K, however the feature seen in ZFC below 10 K is not observed in FC, as the moment is still increasing as $T$ approaches 5 K. It must be noticed here that there is a very clear bifurcation in ZFC and FC curves above 150 K (i.e., $T_N$), which indicates magnetic anisotropy in this system which appears well above $T_N$. The ZFC-FC bifurcation also exhibits local clustering of the spins [16] or anti/ferromagnetic domain growth [17]. Therefore, in the ZFC curve the peak around 10 K can be interpreted in terms of spin-glass transition or freezing of domain walls motion [18]. The ZFC-FC curves merge around 380 K. It may be recalled here that, the ferroelectric Curie temperature ($T_C$) is reported to be around 370 – 380 K [16]. The ferroelectric phase transition might influence the magnetic susceptibility, which results in a magnetic anomaly at 380 K. Our susceptibility measurement results are in good agreement with the reported results [16].

The magnetization as a function of field, measured at $T = 5$ K is shown as an inset to Fig. 3. The *S*-shape of the *M(H)* loops with a very minor hysteresis around origin, gives a clear signature of antiferromagnetic ordering however with spin-glass like anomalies at this temperature. The saturation magnetization ($M_S$), remnant magnetization ($M_R$) and coercive field are found to be 0.12, 9.19 x $10^{-4}$ $\mu_B$/f.u. and 300 Oe, respectively.

PFN is a disordered perovskite, in which $Fe^{+3}$ and $Nb^{+5}$ cations are randomly distributed on the octahedral B sites surrounded by six $O^{-2}$ ions. A super exchange interaction in the disordered regions through $Fe^{+3}$–O–$Nb^{+5}$ is expected to yield AFM ordering below the Néel temperature of 150 K.

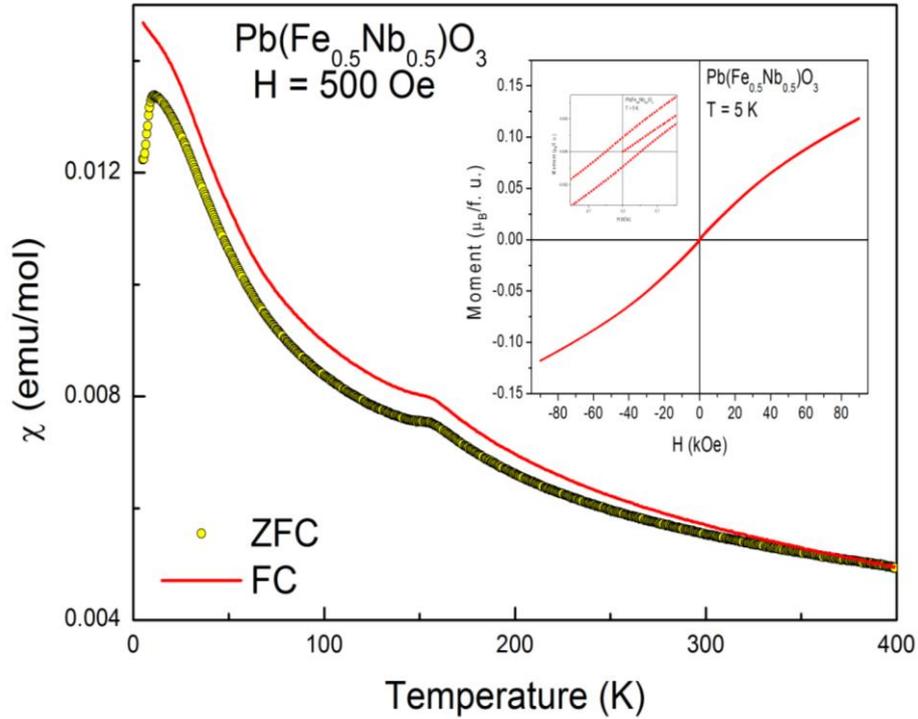

**Figure 3.** Temperature dependent Molar susceptibility (ZFC and FC) for PFN at 500 Oe. Inset shows the *M-H* curve at 5 K for PFN.

*3.3 Neutron diffraction studies*

Neutron Diffraction (ND) patterns of polycrystalline PFN recorded at various temperatures between 2 and 300 K, spanning the $T_N$ exhibits no structural transition across $T_N$, as the structure remains monoclinic (space group *Cm*) in the entire temperature range of measurement. The Rietveld refinement of room temperature ND data refined with *Cm* space group is reported elsewhere [14]. Figure 4 Shows the Rietveld refined ND data taken at 2 K, refinement done for both nuclear and magnetic structure.

The ND data showed that PFN occurs in the monoclinic structure (*Cm* space group) with Fe and Nb atoms occupying 3a sites randomly. In the analysis of structural part (nuclear structure) no extra peaks or splitting of main reflections were observed. The Rietveld analysis of 300 K data was carried out with monoclinic structure (space group *Cm*), based on the reported model [3, 19]. Refinement of the ND data collected at 2 K was also carried out using the same structural model. At $T < T_N$, a new Bragg peak develops around $2\theta = 18.6°$, which can be labeled as the magnetic Bragg peak and which is due to an ordered magnetic structure. In order to determine the magnetic structure, a two-phase refinement of the ND data taken at 2 K was performed.

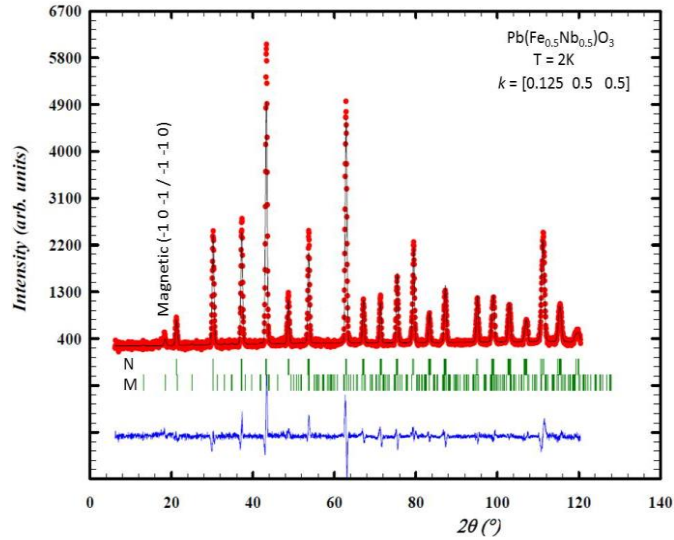

**Figure 4**. Rietveld refinement of the neutron diffraction pattern for PFN taken at 2 K. The first row of vertical tick marks indicate Bragg peaks corresponding to the nuclear (crystallographic) structure. The second row of vertical tick marks indicates magnetic Bragg peak positions.

The magnetic structure can be described in terms of a single propagation vector $k$ = (0.125, 0.5, 0.5). For refinement, basis vectors of the irreducible representations were obtained using *BasIreps* program with in the FullProf suite [20]. Using these basis vectors and the above mentioned propagation vector, the refinement for magnetic structure was carried out assuming Fe is the only magnetic ion in PFN. The magnetic structure could be refined with a G-type antiferromagnetic ordering with the moment constrained along the c-axis. The refined value of the magnetic moment for the Fe cations at 2 K is 5.493 $\mu_B$ per Fe, which is slightly smaller than the spin-only moment of $Fe^{3+}$ ($\mu_{eff} = 2\sqrt{S(S+1)} = 5.9\mu_B$). This deviation suggests that the simple antiferromagnetic model adopted is only an approximation. The atomic coordinates and isotropic temperature factors obtained from the refinement of 2 K ND data are given in Table 1. Table 2 shows the reliability factors from the refinement. The higher $R_{mag}$ value obtained may be due to the another co-existing magnetic phase as is reflected in terms of the broadened magnetic peak.

**Table 1** Structural parameters obtained from Rietveld refinement of ND data of PFN at 2 K. The refined unit cell parameters are $a = 5.6767(6)$ Å, $b = 5.6625(6)$ Å, $c = 4.0198(3)$ Å, $\alpha = \gamma = 90.00°$, $\beta = 89.78(1)°$, Volume = 129.22 (2) Å$^3$.

| Atom | x | y | z | $B_{iso}$ (Å$^2$) | Occ |
|---|---|---|---|---|---|
| **Pb** | 0.000 | 0.000 | 0.000 | 2.085 (171) | 0.500 |
| **Fe/Nb** | 0.5021(43) | 0.000 | 0.4524(40) | 0.357(129) | 0.250 |
| **O$_1$** | 0.4550(47) | 0.000 | -0.0616(55) | 0.655(294) | 0.500 |
| **O$_2$** | 0.2416(47) | 0.2475(29) | 0.4251(30) | 0.242(151) | 1.000 |

The Pb ions reside at the corners of the unit cell, while oxygen octahedra surround the Fe and Nb sites. Although the spin Hamiltonian of PFN has not been established, it is believed that nearest-neighbor Heisenberg interactions are frustrated by next-nearest-neighbor antiferromagnetic ones [21]. The magnetic $Fe^{3+}$ and $Nb^{5+}$ are randomly distributed over the B sites of the perovskite lattice. The nonmagnetic sites locally relieve the geometric frustration of interactions in a spatially random manner. Antiferromagnetic long-range order occurs below $T_N \sim 150$ K. The $Fe^{3+}$ moments are arranged in a simple G-type structure [9] as shown by arrows in Fig. 5. One $Fe^{3+}$ spin is surrounded by six anti-parallel spins on the nearest neighbor $Fe^{3+}$ ions. Due to the structural distortion, the arrangement of neighboring spins is in fact not perfectly anti-parallel. The canted spins induce a weak magnetic moment that couples with the ferroelectric polarization.

**Table 2**. Reliability factors obtained from Rietveld refinement of PFN ND data at 2K.

| Structural parameters | Nuclear structure | Magnetic structure |
|---|---|---|
| Bragg R factor | 7.58 | 9.30 |
| Rf – Factor | 3.42 | 4.44 |
| $R_p$ | 8.24 | 9.20 |
| $R_{wp}$ | 10.7 | 12.0 |
| $R_{exp}$ | 4.60 | 4.62 |
| $\chi^2$ | 5.38 | 6.73 |
| Magnetic R factor | - | 55.9 |

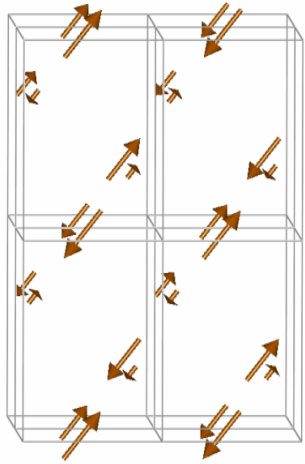

**Figure 5** Canted antiferromagnetic G-type order of $Fe^{3+}$ moments in the collinear phase. The tilt angle Φ randomly varies from site to site averaging to zero.

4. **CONCLUSIONS**

We investigated the magnetic and structural properties of PFN multiferroic. Single phase PFN were synthesized through the low temperature sintering technique i.e. single step solid state reaction method. Two anomalies were found in temperature dependent magnetization. The first was attributed to the PM-to-AFM transition at $T_{N1}$ (150 K). Anomaly at below $T_{N2}$ (10 K) is probably caused by the transition to the spin-glass state, but on the basis of the presented measurements, this conclusion is not sufficient. Further investigation is required to confirm low temperature magnetic phase. The important result of this work consists of low temperature ND studies. The ND measurements at 2 K shows canted G type antiferromagnetic structure. Detailed neutron diffraction studies on PFN as a function of temperature and magnetic field are currently underway and their results will be published separately.

**Acknowledgements**

The work reported here has been carried out under the UGC-DAE CSR collaborative research scheme project number CRS-M-159 of Dr. B. Angadi. SM acknowledges UGC-DAE CSR (Mumbai centre) for financial support in the form of project fellowship.

**REFERENCES**

[1]. G.A. Smolenskii, A. Agranovskaya, S.N. Popov, and V.A. Isupov, Sov. Phys. Tech. Phys. 28 (1958) 2152.

[2]. Nathascia Lampis, Philippe Sciau and Alessandra Geddo Lehmann. Phys.: Condens. Matter 11 (1999) 3489–3500

[3]. S. P. Singh, D. Pandey, S. Yoon, S. Baik, and N. Shin, Appl. Phys. Lett. 90 (2007) 242915.

[4] A. Falqui, N. Lampis, A. Geddo-Lehmann, and G. Pinna, J. Phys. Chem.B 109 (2005) 22967.

[5] A. Kumar, I. Rivera, R. S. Katiyar, and J. F. Scott, Appl. Phys. Lett. 92 (2008) 132913

[6] V.A. Volkov, I.E. Mylnikova, G.A. Smolenskii, Sov. Phys. JETP 15 (1962) 447.

[7] G. A. Smolenskii and I. E. Chupis, Sov. Phys. Usp. 25 (1982) 475

[8] G. L. Platonov, L. A. Drobyshev, Y. Y. Tomashpolskii, Y. N. Venevtsev, Sov. Phys. Crystal. 14 (1970) 692.

[9] S.A. Ivanov, R. Tellgren, H. Rundlof, N.A. Thomas, S. Ananta, J. Phys.: Condens. Matter 12 (2000) 2393.

[10] V. Bonny, M. Bonin, P. Sciau, K.J. Schenk, G. Chapuis, Solid State Commun. 102 (1997) 347.

[11] Smythm D. M, M. P. Harmer, P. Peng, J. Am. Ceram. Soc., 72 (1989) 2276.

[12]   Cho E.S., J. Kim, S. L. Kang, J. J. Appl. Phys., 36 (1997) 5562.

[13]   V. V. Bhat, Basavaraj Angadi and A. M. Umarji, Mater. Sci. and Eng. B. 116 (2) (2005) 131-139.

[14]   Shidaling Matteppanavar, Basavaraj Angadi, Sudhindra Rayaprol., AIP Conference Proceedings 1512 (2013) 1232.

[15]   Rodr´ıguez-Carvajal J  Physica B 192 (1993) 55

[16]   Roman Havlicek, Jana Poltierova Vejpravova, Dariusz Bochenek, J. Phys. Conf. Ser. 200 (2010) 0120581-3

[17]   E. Vincent, V. Dupuis, M. Alba, J. Hammann, and J.-P. Bouchaud, Europhys. Lett. 50 (2000), pp. 674-680.

[18]   H. Chang, Y.-q. Guo, J.-k. Liang, and G.-H. Rao, J. Magn. Magn. Matter. 278 (2004) pp. 306-310.

[19]   Maryanowska A and Pietrzak,  J.  Ferroelectrics 162 (1994) 81

[20]   J. Rodriguez-Carvajal, BASIREPS: a program for calculating irreducible Representations of space groups and basis functions for axial and polar vector properties; 2007

[21]   S. Chillal, M. Thede[2], F. J. Litterst, S. N. Gvasaliya, T. A. Shaplygina, S.G. Lushnikov and A. Zheludev, Phys. Rev. B 87 (2013) 220403(R).